\documentclass[aps,prd,showpacs,amssymb,nofootinbib,preprintnumbers]{revtex4}
\usepackage{slashed}
\usepackage{subfigure}
\usepackage{amsfonts,graphicx,bm}
\usepackage{rotating}
\usepackage{multirow,tabularx}

\begin{document}

\title{Dependence of the Quark-Lepton Complementarity on Parametrizations of the CKM and PMNS Matrices}

\author {Ya-juan Zheng}
\email{yjzheng@mail.sdu.edu.cn}
\affiliation{Department of Physics, Shandong University, Jinan, Shandong 250100, China}
%\date{\today}
\begin{abstract}
The quark-lepton complementarity (QLC) is very suggestive in understanding possible relations between quark and lepton mixing matrices. We explore the QLC relations in all the possible angle-phase parametrizations and point out that they can approximately hold in five parametrizations. Furthermore, the vanishing of the smallest mixing angles in the CKM and PMNS matrices can make sure that the QLC relations exactly hold in those five parametrizations. Finally, the sensitivity of the QLC relations to radiative corrections is also discussed.
\end{abstract}
\maketitle
\section{Introduction}
The success of the Standard Model (SM) in describing the mass origin of elementary particles has satisfied many theoretical physicists but it is now challenged by the existence of neutrino oscillations observed in the solar\cite{SNO}, atmospheric\cite{SK}, reactor\cite{KM} and accelerator\cite{K2K} neutrino experiments, which provide us with convincing evidence for neutrino masses and lepton flavor mixing. The underlying nature of neutrino mixings as compared with that of the quark mixings has inspired a large amount of speculation regarding symmetries in the quark-lepton world as well as other kinds of new physics beyond the SM\cite{King:2003jb}.

In the Pontecorvo-Maki-Nakagawa-Sakata (PMNS)\cite{MNS} lepton mixing matrix, the most distinct feature is the existence of two large mixing angles, which is quite different from the pattern the Cabibbo-Kobayashi-Maskawa (CKM) \cite{CKM} quark mixing matrix. To be specific, the PMNS matrix consists of a large and nearly maximal angle $\vartheta_{23}$ (atmospheric angle), a large but non-maximal angle $\vartheta_{12}$ (solar angle), and a small angle $\vartheta_{13}$ (reactor angle) in the Standard Parametrization. An interesting phenomenological relation between the lepton and quark mixing angles, the so-called quark lepton complementarity (QLC), has been noticed recently\cite{Propose_smirnov}. Namely, the sums of the mixing angles of quarks and leptons for the 1-2 and 2-3 mixings agree with $45^\circ$:
\begin{eqnarray}
\theta_{12}+\vartheta_{12}\simeq45^\circ,~~~~~\theta_{23}+\vartheta_{23}\simeq45^\circ,
\end{eqnarray}
where $\theta_{12}$ and $\theta_{23}$ are quark mixing angles. As for the 1-3 mixing angles of quarks and leptons, a similar relation $\theta_{13}+\vartheta_{13}\simeq45^\circ$ does not hold because their sum is
 less than $10^\circ$.

Attempts to understand the deep meaning behind the QLC relations have been made. It has been interpreted as an evidence for certain quark-lepton symmetry or quark-lepton unification\cite{Propose_Raidal}. Some other aspects of the QLC relations have also been discussed, such as their phenomenological implications\cite{phenomenology} and renormalization group (RG) effects\cite{Smirnov_RGE}. There are also the extended QLC relations proposed and discussed in the Seesaw Mechanisms\cite{extended_QLC}. Recent reviews about the QLC relations can be found in Ref.\cite{Review}. However, whether this relation is accident or not remains an open question. In Ref.\cite{Jarlskog} Jarlskog points out that the QLC relations are parametrization-variant and the specific models are far from being sufficiently pinned-down to be useful for connecting quark and lepton mixing angles like this.

In this paper, we intend to analyze the parametrization dependence of the QLC relations by calculating the mixing angles of each possible parametrization. Among nine angle-phase parametrizations of the CKM and PMNS matrices, we find that five of them can have the approximate QLC relations. If the QLC relations are assumed to exactly hold in a certain parametrization such as the Standard Parametrization, we examine whether they are possible to exactly hold in other parametrizations. Furthermore, the stability of the QLC relations under the RG running is also studied in the Fritzsch-Xing (FX) Parametrization\cite{FXpara}.

The remaining part of this paper is organized as follows. In Section~\ref{section:check}, with the latest experimental data for the CKM and PMNS matrices, we calculate the mixing angles and sum them up in each parametrization to check the QLC relations. Section~\ref{section:transformation} is devoted to examining the relationships between different parametrizations, especially whether the QLC relations in one parametrization can also hold in other parametrizations, and what conditions should be satisfied. The RG running effects on the QLC relations are discussed in the FX Parametrization both in the SM and the Minimal Supersymmetric Standard Model (MSSM) in Section~\ref{section:RGE}.

\section{QLC relations in different angle-phase Parametrizations}\label{section:check}
%\small
%\large
In this section, we try to make numerical calculations of the mixing angles of quark and lepton flavor mixing matrices and examine the QLC relations for all the possible angle-phase parametrizations.

The 3$\times$ 3 CKM quark mixing matrix can be expressed in terms of four independent parameters, which are usually taken as three rotation angles and one CP-violating phase angle. For a clear classification of this kind of angle-phase parametrizations, see \cite{Fritzsch-Xing}. It is pointed out that the CKM matrix $V$, if real and orthogonal, can in general be written as a product of three matrices $R_{12}$, $R_{23}$ and $R_{31}$, which describe simple rotations in the (1, 2), (2, 3) and (3, 1) planes.
\begin{eqnarray}
R_{12}(\theta)=\left(
\begin{array}{ccc}
c_\theta & s_\theta & 0 \\
-s_\theta & c_\theta & 0 \\
0 & 0 & 1
\end{array}\right),
~~R_{23}(\sigma)=\left(
\begin{array}{ccc}
1 & 0 & 0 \\
0 & c_{\sigma} & s_{\sigma} \\
0 & -s_{\sigma} & c_{\sigma}
\end{array}\right),
~~R_{31}(\tau)=\left(
\begin{array}{ccc}
c_\tau & 0 & s_\tau \\
0 & 1 & 0 \\
-s_\tau & 0 & c_\tau
\end{array}\right),
\end{eqnarray}
 where $s_\theta\equiv{\rm sin}\theta,c_\theta\equiv{\rm cos}\theta$, etc. After introducing the CP-violating phase $\phi$, among all the twelve possible products only nine of them are structurally different, as the remaining three products are correlated with each other and leading essentially to the same form. And the specific forms of the nine possible angle-phase parametrizations are listed in the left column of Table~\ref{9-Parameterizations} as P1-P9, and generally P1 corresponds to the Standard Parametrization \cite{PDG} and P2 to the FX Parametrization. For Majorana neutrinos, two additional parameters are needed in PMNS lepton mixing matrix, namely two Majorana CP-violating phase angles, which do not affect oscillations\cite{Strumia}.

In the calculation of quark mixing angles, we take the Wolfenstein parametrization with the accuracy of $O(\lambda^{6})$ as proposed in \cite{Wolfenstein} which is shown as below:
\begin{equation}
V_{\rm CKM} = \left (
\begin{array}{ccc}
1-\frac{1}{2}\lambda^{2}-\frac{1}{8}\lambda^{4}	& \lambda	& A\lambda^{3}(\rho
-i\eta) \\
-\lambda \left [1+ \frac{1}{2}A^{2}\lambda^{4}(2\rho -1) + i
A^{2}\lambda^{4}\eta \right ]	&
1-\frac{1}{2}\lambda^{2}-\frac{1}{8}\left (4A^{2}+1\right)\lambda^{4} 	&
A\lambda^{2} \\
A\lambda^{3}(1-\rho-i\eta)	& -A\lambda^{2}\left [1+\frac{1}{2}\lambda^{2}(2\rho
-1 )
+i\lambda^{2}\eta \right ] 	& 1-\frac{1}{2}A^{2}\lambda^{4}
\end{array}
\right ) \;
%		(4)
\end{equation}
To calculate the moduli of the mixing matrix elements, we adopt the following inputs given by the Particle Data Group \cite{PDG}:
\begin{eqnarray}
\lambda = 0.2257^{+0.0009}_{-0.0010},~~~~A=0.814^{+0.021}_{-0.022},~~~~\bar{\rho}=0.135^{+0.031}_{-0.016},~~~~~~\bar{\eta}=0.349^{+0.015}_{-0.017}
\end{eqnarray}
where
\begin{eqnarray}
\nonumber
&\bar{\rho}&=\rho - \frac{1}{2}\rho \lambda^2 +\left(\frac{1}{2}A^2\rho -\frac{1}{8} \rho-A^2(\rho^2-\eta^2)\right)\lambda^4+O(\lambda^6), \\
&\bar{\eta} &=\eta-\frac{1}{2}\eta \lambda^2+\left(\frac{1}{2}A^2\eta -\frac{1}{8}\eta-2 A^2\rho\eta\right)\lambda^4+O(\lambda^6).
\end{eqnarray}\begin{tiny}\end{tiny}
Then we obtain
\begin{eqnarray}
|V_{\rm CKM}|=\left(
\begin{array}{ccc}
0.974205^{-0.00021}_{+0.00023} & 0.225700^{+0.00090}_{-0.00100} & 0.003592^{+0.00040}_{-0.00034} \\
0.225560^{+0.00090}_{-0.00100}  & 0.973346^{-0.00027}_{+0.00029} & 0.041466^{+0.00141}_{-0.00148} \\
0.008733^{+0.00011}_{-0.00027} & 0.040709^{+0.00144}_{-0.00148} & 0.999140^{-0.00006}_{+0.00006}
\end{array}\right).
\end{eqnarray}
This result allows us to calculate the mixing angles in all the nine parameterizations according to the relations between angles and moduli.

For lepton mixing angles, the Standard Parameterization is expressed in terms of three mixing angles $\vartheta_{12}$, $\vartheta_{13}$, $\vartheta_{23}$ and one CP-violating phase angle $\varphi$. As shown below, the first row and third column have a pretty simple form.
\begin{eqnarray}
V_{\rm PMNS} = \left (
\begin{array}{ccc}
c^{~}_{12} c_{13}         & s^{~}_{12} c_{13}       & s_{13} \cr
-c^{~}_{12} s_{23} s_{13} - s^{~}_{12} c_{23} e^{-{\rm i}\varphi}
& -s^{~}_{12} s_{23} s_{13} + c^{~}_{12} c_{23} e^{-{\rm i}\varphi}   & s_{23} c_{13} \cr
-c^{~}_{12} c_{23} s_{13} + s^{~}_{12} s_{23} e^{-{\rm i}\varphi}
& -s^{~}_{12} c_{23} s_{13} - c^{~}_{12} s_{23} e^{-{\rm i}\varphi}   & c_{23} c_{13}
\end{array}
\right),
\end{eqnarray}
 where $s_{ij}={\rm sin}\vartheta_{ij}, c_{ij}={\rm cos}\vartheta_{ij}~(i,j=1,2,3)$. With the latest global fit of the experimental data given in\cite{Fogli:2009zza}, the three mixing angles read
\begin{eqnarray}
\nonumber
{\rm sin}^2\vartheta_{12}=0.312(1^{+0.128}_{-0.109})~(2\sigma)\\
\nonumber
{\rm sin}^2\vartheta_{23}=0.466(1^{+0.292}_{-0.215}) ~(2\sigma) \\
{\rm sin}^2\vartheta_{13}=0.016\pm 0.010 ~(1\sigma)
\end{eqnarray}
Due to the smallness of $\vartheta_{13}\simeq~(7.27^{+2.012}_{-2.824})^\circ$, those terms including ${\rm sin}\vartheta_{13}$ could be neglected in the (2, 1), (2, 2), (3, 1) and (3, 2) entries in the Standard Parameterization. Thus the moduli of the mixing matrix elements can be obtained,
\begin{equation}
\left|V_{\rm PMNS}\right| = \left (
\begin{array}{ccc}
0.822795^{-0.0283}_{+0.0244}	& 0.554083^{+0.0314}_{-0.0284}	& 0.126491^{+0.0348}_{-0.0490} \\
0.408176^{-0.0340}_{+0.0117}	&
0.606129^{-0.0983}_{+0.0705} 	&
0.677159^{+0.0886}_{-0.0742} \\
0.381303^{+0.0790}_{-0.0624} &  0.566223^{+0.0584}_{-0.0523}	& 0.724883^{-0.1023}_{+0.1023}
\end{array}
\right ). \;
\end{equation}
 With the relations between the moduli of mixing matrix elements and mixing angles in each of the parametrization, we can get all the mixing angles.

The numerical results of quark and lepton mixing angles as well as their QLC relations are listed in the right column of Table~\ref{9-Parameterizations}. It is obvious from the table that the QLC relations approximately hold in P1, P2, P3, P4 and P5 parametrizations but suffer from large deviation in the remaining four parametrizations. Thus the QLC relations are indeed parameterization-dependent. Furthermore, the distinct feature for those parametrizations accommodating the QLC relations is that they all have a simple form in their (1, 3) entries.

\section{Conditions for the exact QLC relations Transformation}\label{section:transformation}
Since the QLC relations depend on the forms of parametrizations, the exploration of those parametrizations which ensure this relation is necessary and pressing. Based on the hypothesis that the QLC relations hold in the Standard Parametrization,
\begin{eqnarray}
\theta_{12}+\vartheta_{12}=45^\circ,~~~\theta_{23}+\vartheta_{23}=45^\circ,
\label{qlc2}
\end{eqnarray}
 we aim to see under what conditions this is still the case for the corresponding angles in the other parametrizations.
 Take the FX Parametrization (i.e. P2 in Table.~\ref{9-Parameterizations}) as an example, we have
\begin{eqnarray}
{\rm tan}\theta_d=\frac{|V_{td}|}{|V_{ts}|}=\frac{|c_{\theta_{12}} c_{\theta_{23}} s_{\theta_{13}} - s_{\theta_{12}} s_{\theta_{23}}e^{-i\phi}|}{|s_{\theta_{12}} c_{\theta_{23}} s_{\theta_{13}} + c_{\theta_{12}} s_{\theta_{23}}e^{-i\phi}|},~~~~
{\rm cos}\theta=|V_{tb}|=|c_{\theta_{23}} c_{\theta_{13}}|.
\label{relation}
\end{eqnarray}
We first consider $\theta_d+\vartheta_\nu$ which is corresponding to the first relation in Eq.(\ref{qlc2}):
\begin{eqnarray}
\nonumber
{\rm tan}(\theta_d+\vartheta_\nu)&=& \frac{{\rm tan}\theta_d+{\rm tan}\vartheta_\nu}{1-{\rm tan}\theta_d {\rm tan}\vartheta_\nu }\\
&=&\frac{{\cal C} {\cal A}+{\cal D}{\cal B}}{{\cal D}{\cal A}-{\cal C}{\cal B}},
\label{tan}
\end{eqnarray}
where ${\cal A}$, ${\cal B}$, ${\cal C}$ and ${\cal D}$ are defined as
\begin{eqnarray}
\nonumber
{\cal A}&=&|s_{\vartheta_{12}} c_{\vartheta_{23}} s_{\vartheta_{13}}+c_{\vartheta_{12}} s_{\vartheta_{23}}e^{-i\varphi}|,~~~{\cal B}=|c_{\vartheta_{12}} c_{\vartheta_{23}}s_{\vartheta_{13}}+s_{\vartheta_{12}}s_{\vartheta_{23}}e^{-i\varphi}|, \\
{\cal C}&=&|c_{\theta_{12}} c_{\theta_{23}} s_{\theta_{13}}-s_{\theta_{12}}s_{\theta_{23}}e^{-i\phi}|,~~~~{\cal D}=|s_{\theta_{12}}c_{\theta_{23}}s_{\theta_{13}}+c_{\theta_{12}}s_{\theta_{23}}e^{-i\phi}|.
\label{ABCD}
\end{eqnarray}
Using the QLC relation in Eq.(\ref{qlc2}), one can get ${\cal A}$ and ${\cal B}$ expressed in the form of the Standard Parametrization:
\begin{eqnarray}
\nonumber
{\cal A}&=&\frac{1}{2}|(c_{\theta_{12}}-s_{\theta_{12}})(c_{\theta_{23}}+s_{\theta_{23}})s_{\vartheta_{13}}+(c_{\theta_{12}}+s_{\theta_{12}})(c_{\theta_{23}}-s_{\theta_{23}})e^{-i\varphi}|, \\
{\cal B}&=&\frac{1}{2}|(c_{\theta_{12}}+s_{\theta_{12}})(c_{\theta_{23}}+s_{\theta_{23}})s_{\vartheta_{13}}+(c_{\theta_{12}}-s_{\theta_{12}})(c_{\theta_{23}}-s_{\theta_{23}})e^{-i\varphi}|.
\end{eqnarray}
After substituting the above expressions of ${\cal A}$ and ${\cal B}$ into Eq.({\ref{tan}}), we find that it is hard to deduce any useful conclusion from it. Furthermore, we assume that the corresponding smallest angles in the Standard Parametrization for quark and lepton mixings are vanishing, i.e. $\theta_{13}=\vartheta_{13}=0^\circ$, thus
\begin{eqnarray}
\nonumber
{\cal A}&=&\frac{1}{2}|(c_{\theta_{12}}+s_{\theta_{12}})(c_{\theta_{23}}-s_{\theta_{23}})|,~~
{\cal B}=\frac{1}{2}|(c_{\theta_{12}}-s_{\theta_{12}})(c_{\theta_{23}}-s_{\theta_{23}})|, \\
{\cal C}&=&|s_{\theta_{12}}s_{\theta_{23}}|,~~~~~~~~~~~~~~~~~~~~~~~~~~
{\cal D}=|c_{\theta_{12}}s_{\theta_{23}}|,
\end{eqnarray}
which leads to a remarkable result,
\begin{eqnarray}
\nonumber
{\rm tan}(\theta_d+\vartheta_\nu)&=&\frac{{\cal C} {\cal A}+{\cal D}{\cal B}}{{\cal D}{\cal A}-{\cal C}{\cal B}} \\
\nonumber
&=&\frac{|s_{\theta_{12}}s_{\theta_{23}}| |(c_{\theta_{12}}+s_{\theta_{12}})(c_{\theta_{23}}-s_{\theta_{23}})|+|c_{\theta_{12}}s_{\theta_{23}}||(c_{\theta_{12}}-s_{\theta_{12}})(c_{\theta_{23}}-s_{\theta_{23}})|}{|c_{\theta_{12}}s_{\theta_{23}}||(c_{\theta_{12}}+s_{\theta_{12}})(c_{\theta_{23}}-s_{\theta_{23}})|-|s_{\theta_{12}}s_{\theta_{23}}||(c_{\theta_{12}}-s_{\theta_{12}})(c_{\theta_{23}}-s_{\theta_{23}})|} \\
\nonumber
&=&\left|\frac{s_{\theta_{12}}(c_{\theta_{12}}+s_{\theta_{12}})+c_{\theta_{12}}(c_{\theta_{12}}-s_{\theta_{12}})}{c_{\theta_{12}}(c_{\theta_{12}}+s_{\theta_{12}})-s_{\theta_{12}}(c_{\theta_{12}}-s_{\theta_{12}})}\right| \\
&=&1,
\end{eqnarray}
from which the QLC relation in the FX Parametrization $\theta_d+\vartheta_\nu=45^\circ$ exactly holds.

 Now we turn to consider the second relation of Eq.(\ref{relation}),
\begin{eqnarray}
\nonumber
{\rm cos}(\theta+\vartheta) &=& c_{\theta} c_{\vartheta} -s_{\theta} s_{\vartheta}\\
\nonumber
&=&c_{\theta_{23}}c_{\theta_{13}}c_{\vartheta_{23}}c_{\vartheta_{13}}-\sqrt{(1-c_{\theta_{23}}^2c_{\theta_{13}}^2)(1-c_{\vartheta_{23}}^2c_{\vartheta_{13}}^2)} \\
\nonumber
&=&\frac{\sqrt{2}}{2}c_{\theta_{23}}c_{\theta_{13}}c_{\vartheta_{13}}(c_{\theta_{23}}+s_{\theta_{23}})-\sqrt{(1-c_{\theta_{23}}^2c_{\theta_{13}}^2)\left[1-\frac{1}{2}(c_{\theta_{23}}+s_{\theta_{23}})^2c_{\vartheta_{13}}^2\right]}.
\end{eqnarray}
With the help of the QLC relation for $\theta_{23}$ and $\vartheta_{23}$ and the assumption of the vanishing smallest mixing angles $\theta_{13}=\vartheta_{13}=0^\circ$, one can obtain
\begin{eqnarray}
\nonumber
{\rm cos}(\theta+\vartheta) &=& \frac{\sqrt{2}}{2}c_{\theta_{23}}(c_{\theta_{23}}+s_{\theta_{23}})-\sqrt{s_{\theta_{23}}^2\left[1-\frac{1}{2}(1-2c_{\theta_{23}}s_{\theta_{23}})\right]} \\
\nonumber
&=&\frac{\sqrt{2}}{2}c_{\theta_{23}}(c_{\theta_{23}}+s_{\theta_{23}})-\frac{\sqrt{2}}{2}s_{\theta_{23}}(c_{\theta_{23}}-s_{\theta_{23}})\\
\nonumber
&=&\frac{\sqrt{2}}{2}.
\end{eqnarray}
Again, the QLC relation holds for the FX Parametrization in this situation. Namely,
\begin{eqnarray}
\theta_d+\vartheta_\nu=45^\circ~~ {\rm and}~~\theta+\vartheta=45^\circ.
\end{eqnarray}

In fact, under the condition of $\theta_{13}=\vartheta_{13}=0^\circ$, the conclusion that the QLC relations hold in these parametrizatons can be exactly obtained. For example, from Eq.(\ref{relation}) we can easily get ${\rm tan}\theta_d=|{\rm tan}\theta_{12}|$ and ${\rm cos}\theta=|{\rm cos}\theta_{23}|$ in the FX Parametrization if $\theta_{13}=\vartheta_{13}=0^\circ$. So is the case in the lepton sector, i.e. ${\rm tan}\vartheta_\nu=|{\rm tan}\vartheta_{12}|$ and ${\rm cos}\vartheta=|{\rm cos}\vartheta_{23}|$. Hence, the QLC relations hold in the FX Parametrization. And the same conclusion can also be obtained for P3, P4 and P5 Parametrizations in a similar procedure. The reason is simple that these five parametrizations are essentially equivalent to one another in the $\theta_{13}=\vartheta_{13}=0^\circ$ limit.
\section{on the stability of QLC relations RG running}\label{section:RGE}
As proposed in many papers, the quark-lepton symmetry implied by the QLC relations means that physics responsible for these relations should be realized at some scales which might be the quark-lepton unification scale, $\Lambda_{\rm GUT}$ or even higher scales, and the RG effects has been discussed in the framework of the Standard Parametrization\cite{Smirnov_RGE,Antusch:2003kp}. Since there are specific advantages in the FX Parametrization for the study of fermion mass matrices and B-meson physics\cite{Fritzsch-Xing}, it is useful to examine the sensitivity of the QLC relations to the RG effects in this parametrization. And it has been shown that the RG equations of quark and lepton mixing angles have a particularly simple form in the FX Parametrization\cite{XRGE_q,XRGE_l}. Assume that QLC relations hold exactly at the scale $M_Z$ in this parametrization:
\begin{eqnarray}
\theta_d+\vartheta_\nu = 45^\circ,~~~\theta+\vartheta  =45^\circ,
\label{assumption_3}
\end{eqnarray}
thus
\begin{eqnarray}
\dot{\theta}_d+\dot{\vartheta}_\nu=0,~~~\dot{\theta}+\dot{\vartheta}=0,
\label{qlc_RGE2}
\end{eqnarray}
where $\dot{\theta}=\displaystyle{\frac{{\rm d}\theta}{{\rm d}t}}$ with $t\equiv {\rm ln}~(\mu/M_Z)$. We already have the RG equations of three quark mixing angles \cite{XRGE_q} and three Dirac neutrino mixing angles \cite{XRGE_l} in FX Parametrization:
\begin{eqnarray}
\renewcommand\arraystretch{1.5}
\left\{\begin{array}{l}
\dot{\theta}_u =
\displaystyle{-\frac{1}{32\pi^2} }C y^2_b \sin 2\theta_u \sin^2\theta, \\
\dot{\theta}_d = \displaystyle{-\frac{1}{32\pi^2 }} C y^2_t \sin 2\theta_d \sin^2\theta, \\
\dot{\theta}  =  \displaystyle{-\frac{1}{32\pi^2}} C
\left( y^2_b + y^2_t \right) \sin 2\theta,
\end{array}\right.~~~~~~~~
\left\{\begin{array}{l}
\dot{\vartheta}_l  =  \displaystyle{+ \frac{Cy^2_\tau}{16\pi^2}} ~ c^{}_{\nu}
s^{}_{\nu} c_{\vartheta} c^{}_{\varphi} \left ( \xi^{}_{13} - \xi^{}_{23} \right ), \\
\dot{\vartheta}_\nu  = \displaystyle{ + \frac{Cy^2_\tau}{16\pi^2}} ~ c^{}_{\nu}
s^{}_{\nu} \left [ s_{\vartheta}^2 \xi^{}_{12} + c_{\vartheta}^2 \left ( \xi^{}_{13} -
\xi^{}_{23} \right ) \right ], \\
\dot{\vartheta}  = \displaystyle{ + \frac{Cy^2_\tau}{16\pi^2}} ~ c_{\vartheta} s_{\vartheta} \left (
s^2_{\nu} \xi^{}_{13} + c^2_{\nu} \xi^{}_{23} \right ),
\end{array}\right.
\label{angleRGE}
\end{eqnarray}
where $C=-1.5~(+1)$ in the SM (MSSM),  $\xi^{}_{ij} \equiv \left (y^2_i + y^2_j \right )/\left
(y^2_i - y^2_j \right )$ and $y_\alpha,~y_a $ and $y_i$ $(\alpha=\tau,a=b,t$ and $ i=1,2,3)$ stand respectively for the eigenvalues of the Yukawa coupling matrices of charged leptons, quarks and neutrinos. In the case of the SM, the Yukawa couplings $y_i=\displaystyle{\frac{m_i}{v}}(i=1,2,3)$, where the Higgs vacuum expection value (VEV) $v$ is $174$ GeV. In the MSSM, $m_\alpha=y_\alpha v {\rm sin}\beta,~m_\gamma=y_{\gamma} v {\rm cos}\beta~(\alpha=u,c,t,\gamma=d,s,b,e,\mu,\tau)$, where tan$\beta$ is the ratio of two Higgs VEV's.

Some qualitative comments on the main features of Eq.(\ref{angleRGE}) are in order.

(a) For the RG equations of quark flavor mixing angles in both the SM and MSSM, noticing that the value of $\theta$ is very small, we can safely claim that the RG running effects of $\theta_u$, $\theta_d$ and $\theta$ are highly suppressed. As a result, three quark mixing angles in FX Parametrization will not change a lot under the RG running.

(b)  In the lepton sector in the SM case, the derivatives of three mixing angles are proportional to $y_\tau^2=\displaystyle{\left(\frac{m_\tau}{v}\right)^2}\simeq10^{-4}$\cite{Xing_running}. Notice that $\xi_{ij}=-\displaystyle{\frac{m_i^2+m_j^2}{\Delta m_{ji}^2}}$, with $\Delta m_{ji}^2=m_j^2-m_i^2$, $\Delta m_{21}^2\simeq7.7\times10^{-5}~{\rm eV^2}$ and $|\Delta m_{32}^2|\approx|\Delta m_{31}^2|\approx 2.4\times 10^{-3}~{\rm eV^2} $\cite{Fogli:2009zza}. Thus the most sensitive angle to radiative corrections is $\vartheta_\nu$ whose RG equation is the only one that consists of $\xi_{12}$. But we still cannot expect large running effects on $\theta_{\nu}$ because the loop factor $1/16\pi^2$ makes the derivative even smaller. While in the MSSM case, where $y_\tau=\displaystyle{\frac{m_\tau}{v {\rm cos}\beta}}$, the RG running effects could be enhanced when ${\rm tan}\beta$ is significantly large.

As a result, if we sum up the derivatives of the corresponding mixing angles of quark and lepton sectors in Eq.(\ref{angleRGE}), we can conclude that in the SM case, the QLC relations are essentially stable under the RG running and in the MSSM case, these relations might become unstable only when ${\rm tan}\beta$ is sufficiently large.
\section{summary}\label{section:summary}
 To understand the deep meaning of the quark and lepton world, the quark-lepton symmetry topic has drawn a lot of attention recent years. Among many of the aspects that imply the symmetry and unification in quark and lepton sectors, the QLC relations between the mixing angles of the CKM and PMNS matrices have been considered very interesting and suggestive.
In this paper, we have calculated the QLC relations for each of the angle-phase parametrizations and find that these relations are parametrization-dependent. Furthermore, the distinct feature of those parametrizations which can approximately accommodate the QLC relations is that they all have a simple form in the (1, 3) entries. Then based on the assumption that the QLC relations hold exactly in the Standard Parametrization we make an exploration in the FX Parametrization and get the conclusion that these relations can also hold as long as the smallest mixing angle $\theta_{13}$ is vanishing. Finally, we make clear that the QLC relations can essentially stay stable under the RG running effects in the SM and MSSM unless the value of ${\rm tan}\beta$ is sufficiently large.
%\section{Acknowledgements}
%\begin{acknowledgements}
\begin{acknowledgements}
The author is greatly indebted to Professor Z.Z. Xing for stimulating discussions, constant instruction and polishing the manuscript with great care, and to S. Luo for warm hospitality during her visiting stay in Beijing IHEP where this work was done. She is also grateful to W.T. Deng, Y.K. Song, H. Zhang and Professor Z.G. Si for useful discussions and comments. This work is supported in part by NSFC and Natural Science Foundation of Shandong Province.
\end{acknowledgements}
%%%%%%%%%%%%%%%%%% Table 1 %%%%%%%%%%%%%%%%%%%%
\small
\begin{table}\label{table}
\caption{Classification of different parametrizations for the flavor mixing
matrix and the QLC relations.}
\vspace{-1cm}
\begin{center}
\begin{tabular}{ccc} \\ \hline\hline
Parametrization &~& Quark Lepton Complementarity
\\  \hline
\underline{{\it P1:} ~ $V \; = \; R_{23}(\theta_{23}) ~ R_{31}(\theta_{13}, \phi)
~ R_{12}(\theta_{12})$}
&& $\theta_{12}/\theta_{23}/\theta_{13}~~~~~~~~\vartheta_{12}/\vartheta_{23}/\vartheta_{13}~~~~~~~~~~~~~~~~~~~~~~$
\\
$\left ( \matrix{
c^{~}_{12} c_{13}         & s^{~}_{12} c_{13}       & s_{13} \cr
-c^{~}_{12} s_{23} s_{13} - s^{~}_{12} c_{23} e^{-{\rm i}\phi}
& -s^{~}_{12} s_{23} s_{13} + c^{~}_{12} c_{23} e^{-{\rm i}\phi}   & s_{23} c_{13} \cr
-c^{~}_{12} c_{23} s_{13} + s^{~}_{12} s_{23} e^{-{\rm i}\phi}
& -s^{~}_{12} c_{23} s_{13} - c^{~}_{12} s_{23} e^{-{\rm i}\phi}   & c_{23} c_{13} \cr} \right )
$
&& $\matrix{
(13.04^{+0.053}_{-0.059})^\circ+ (33.96^{+2.430}_{-2.137})^\circ=(47.00^{+2.483}_{-2.196})^\circ \cr
(2.37^{+0.081}_{-0.085})^\circ+(43.05^{+7.839}_{-5.834})^\circ=(45.42^{+7.920}_{-5.919})^\circ \cr
(0.20^{+0.023}_{-0.020})^\circ+(7.27^{+2.012}_{-2.824})^\circ=(7.47^{+2.035}_{-2.844})^\circ \cr} $ \\ \\
%-----------------------------------------------------------------
\underline{{\it P2:} ~ $V \; = \; R_{12}(\theta_u) ~ R_{23}( \theta, \phi)
~ R^{-1}_{12}(\theta_d)$}
&&$\theta_u/\theta_d/\theta~~~~~~~~~~\vartheta_l/\vartheta_\nu/\vartheta~~~~~~~~~~~~~~~~~~~~~~  $\\
$\left ( \matrix{
s^{~}_{u} s^{~}_{d} c_{ } + c^{~}_{u} c^{~}_{d} e^{-{\rm i}\phi}  &
s^{~}_{u} c^{~}_{d} c_{ } - c^{~}_{u}
s^{~}_{d} e^{-{\rm i}\phi}     & s^{~}_{u} s_{ } \cr
c^{~}_{u} s^{~}_{d} c_{ } - s^{~}_{u} c^{~}_{d} e^{-{\rm i}\phi}  &
c^{~}_{u} c^{~}_{d} c_{ } + s^{~}_{u}
s^{~}_{d} e^{-{\rm i}\phi}     & c^{~}_{u} s_{ } \cr
- s^{~}_{d} s_{ }    & - c^{~}_{d}
s_{ }      & c_{ } \cr} \right )
$
&& $\matrix{
 (4.95^{+0.363}_{-0.305})^\circ+(10.58^{+1.310}_{-3.261})^\circ=(15.53^{+1.637}_{-3.566})^\circ  \cr
(12.11^{-0.262}_{+0.065})^\circ+(33.96^{+2.430}_{-2.137})^\circ=(46.67^{+2.168}_{-2.072})^\circ  \cr
(2.38^{+0.081}_{-0.085})^\circ+(43.54^{+7.956}_{-6.099})^\circ=(45.92^{+8.037}_{-6.184})^\circ  \cr}$ \\ \\
%-----------------------------------------------------------------
\underline{{\it P3:} ~ $V \; = \; R_{23}(\theta_d) ~ R_{12}(\theta, \phi)
~ R^{-1}_{23}(\theta_u)$}
&& $\theta_u/\theta_d/\theta~~~~~~~~~~\vartheta_l/\vartheta_\nu/\vartheta~~~~~~~~~~~~~~~~~~~~~~  $ \\
$\left ( \matrix{
c^{~}_{\theta}  & s^{~}_{\theta} c_{u}    & -s^{~}_{\theta} s_{u} \cr
-s^{~}_{\theta} c_{d}      & c^{~}_{\theta} c_{d} c_{u} + s_{d} s_{u} e^{-{\rm i}\phi}
& -c^{~}_{\theta} c_{d} s_{u} + s_{d} c_{u} e^{-{\rm i}\phi} \cr
s^{~}_{\theta} s_{d}       & -c^{~}_{\theta} s_{d} c_{u} + c_{d} s_{u} e^{-{\rm i}\phi}
& c^{~}_{\theta} s_{d} s_{u} + c_{d} c_{u} e^{-{\rm i}\phi} \cr} \right )
$
&& $\matrix{
(0.91^{+0.096}_{-0.083})^\circ+(12.86^{+2.538}_{-4.477})^\circ=(13.77^{+2.634}_{-4.560})^\circ \cr
(2.22^{+0.019}_{-0.059})^\circ+(43.05^{+7.839}_{-5.834})^\circ=(45.27^{+7.858}_{-5.893})^\circ\cr
(13.04^{+0.053}_{-0.059})^\circ+(34.63^{+2.758}_{-2.538})^\circ=(47.67^{+2.811}_{-2.597})^\circ \cr} $ \\ \\
%-----------------------------------------------------------------
\underline{{\it P4:} ~ $V \; = \; R_{23}(\sigma) ~ R_{12}(\theta, \phi)
~ R^{-1}_{31}(\rho)$}
&& $\theta_u/\theta_d/\theta~~~~~~~~~~\vartheta_l/\vartheta_\nu/\vartheta~~~~~~~~~~~~~~~~~~~~~~  $
\\
$\left ( \matrix{
c^{~}_{\theta} c_{u}         & s^{~}_{\theta}        & -c^{~}_{\theta} s_{u} \cr
-s^{~}_{\theta} c_{d} c_{u} + s_{d} s_{u} e^{-{\rm i}\phi}   & c^{~}_{\theta} c_{d}
& s^{~}_{\theta} c_{d} s_{u} + s_{d} c_{u} e^{-{\rm i}\phi} \cr
s^{~}_{\theta} s_{d} c_{u} + c_{d} s_{u} e^{-{\rm i}\phi}    & -c^{~}_{\theta} s_{d}
& -s^{~}_{\theta} s_{d} s_{u} + c_{d} c_{u} e^{-{\rm i}\phi} \cr} \right )
$
&& $\matrix{
(0.21^{+0.023}_{-0.020})^\circ+(8.74^{+2.733}_{-3.516})^\circ=(8.95^{+2.756}_{-3.536})^\circ\cr
(2.39^{+0.086}_{-0.088})^\circ+(43.05^{+7.839}_{-5.834})^\circ=(45.44^{+7.925}_{-5.922})^\circ \cr
(13.04^{+0.053}_{-0.059})^\circ+(33.65^{+2.189}_{-1.935})^\circ=(46.69^{+2.242}_{-1.994})^\circ \cr} $ \\ \\
%-----------------------------------------------------------------
\underline{{\it P5:} ~ $V \; = \; R_{31}(\rho) ~ R_{23}(\sigma, \phi)
~ R^{-1}_{12}(\theta)$}
&&  $\theta_u/\theta_d/\theta~~~~~~~~~~\vartheta_l/\vartheta_\nu/\vartheta~~~~~~~~~~~~~~~~~~~~~~  $
\\
$\left ( \matrix{
-s^{~}_{\theta} s_{d} s_{u} + c^{~}_{\theta} c_{u} e^{-{\rm i}\phi}
& -c^{~}_{\theta} s_{d} s_{u} - s^{~}_{\theta} c_{u} e^{-{\rm i}\phi}     & c_{d} s_{u} \cr
s^{~}_{\theta} c_{d}       & c^{~}_{\theta} c_{d}     & s_{d} \cr
-s^{~}_{\theta} s_{d} c_{u} - c^{~}_{\theta} s_{u} e^{-{\rm i}\phi}
& -c^{~}_{\theta} s_{d} c_{u} + s^{~}_{\theta} s_{u} e^{-{\rm i}\phi}     & c_{d} c_{u} \cr} \right )
$
&& $\matrix{
(0.21^{+0.023}_{-0.020})^\circ+(9.90^{+4.622}_{-4.326})^\circ=(10.11^{+4.645}_{-4.346})^\circ \cr
(2.38^{+0.081}_{-0.085})^\circ+(42.62^{+7.354}_{-5.537})^\circ=(45.00^{+7.435}_{-5.622})^\circ \cr
(13.05^{+0.054}_{-0.059})^\circ+(33.96^{+2.430}_{-2.137})^\circ=(47.01^{+2.484}_{-2.196})^\circ \cr} $ \\ \\
%-----------------------------------------------------------------
\underline{{\it P6:} ~ $V \; = \; R_{12}(\theta) ~ R_{31}(\theta_u, \phi)
~ R^{-1}_{23}(\theta_d)$}
&& $\theta_u/\theta_d/\theta~~~~~~~~~~\vartheta_l/\vartheta_\nu/\vartheta~~~~~~~~~~~~~~~~~~~~~~  $
\\
$\left ( \matrix{
c^{~}_{\theta} c_{u}         & c^{~}_{\theta} s_{d} s_{u} + s^{~}_{\theta} c_{d} e^{-{\rm i}\phi}
& c^{~}_{\theta} c_{d} s_{u} - s^{~}_{\theta} s_{d} e^{-{\rm i}\phi} \cr
-s^{~}_{\theta} c_{u}        & -s^{~}_{\theta} s_{d} s_{u} + c^{~}_{\theta} c_{d} e^{-{\rm i}\phi}
& -s^{~}_{\theta} c_{d} s_{u} - c^{~}_{\theta} s_{d} e^{-{\rm i}\phi} \cr
-s_{u}       & s_{d} c_{u}   & c_{d} c_{u} \cr} \right )
$
&& $\matrix{
(0.50^{+0.006}_{-0.015})^\circ+(22.41^{+4.993}_{-3.818})=(22.91^{+4.999}_{-3.833})^\circ \cr
(2.33^{+0.083}_{-0.085})^\circ+(37.99^{+7.102}_{-5.080})^\circ=(40.32^{+7.185}_{-5.165})^\circ \cr
(13.04^{+0.053}_{-0.059})^\circ+(26.39^{-1.164}_{-0.021})^\circ=(39.43^{-1.111}_{-0.080})^\circ \cr} $ \\ \\
%-----------------------------------------------------------------
\underline{{\it P7:} ~ $V \; = \; R_{31}(\theta_u) ~ R_{12}(\theta, \phi)
~ R^{-1}_{31}(\theta_d)$}
&& $\theta_u/\theta_d/\theta~~~~~~~~~~\vartheta_l/\vartheta_\nu/\vartheta~~~~~~~~~~~~~~~~~~~~~~  $
\\
$\left ( \matrix{
c^{~}_{\theta} c_{u} c_{d} + s_{u} s_{d} e^{-{\rm i}\phi}      & s^{~}_{\theta} c_{u}
& -c^{~}_{\theta} c_{u} s_{d} + s_{u} c_{d} e^{-{\rm i}\phi} \cr
-s^{~}_{\theta} c_{d}       & c^{~}_{\theta}        & s^{~}_{\theta} s_{d} \cr
-c^{~}_{\theta} s_{u} c_{d} + c_{u} s_{d} e^{-{\rm i}\phi}     & -s^{~}_{\theta} s_{u}
& c^{~}_{\theta} s_{u} s_{d} + c_{u} c_{d} e^{-{\rm i}\phi} \cr} \right )
$
&& $\matrix{
(10.22^{+0.315}_{-0.320})^\circ+(45.62^{+1.233}_{-1.268})^\circ=(55.84^{+1.537}_{-1.577})^\circ \cr
(10.42^{+0.304}_{-0.320})^\circ+(58.92^{+5.037}_{-3.769})^\circ=(69.34^{+5.341}_{-4.089})^\circ \cr
(13.26^{+0.067}_{-0.073})^\circ+(52.69^{+6.791}_{-5.274})^\circ=(65.95^{+6.858}_{-5.347})^\circ \cr} $ \\ \\
%--------------------------------------------------------------
\underline{{\it P8:} ~ $V \; = \; R_{12}(\theta) ~ R_{23}(\theta_d, \phi)
~ R_{31}(\theta_u)$}
&& $\theta_u/\theta_d/\theta~~~~~~~~~~\vartheta_l/\vartheta_\nu/\vartheta~~~~~~~~~~~~~~~~~~~~~~  $
\\
$\left ( \matrix{
-s^{~}_{\theta} s_{d} s_{u} + c^{~}_{\theta} c_{u} e^{-{\rm i}\phi}       & s^{~}_{\theta} c_{d}     &
s^{~}_{\theta} s_{d} c_{u} + c^{~}_{\theta} s_{u} e^{-{\rm i}\phi} \cr
-c^{~}_{\theta} s_{d} s_{u} - s^{~}_{\theta} c_{u} e^{-{\rm i}\phi}       & c^{~}_{\theta} c_{d}     &
c^{~}_{\theta} s_{d} c_{u} - s^{~}_{\theta} s_{u} e^{-{\rm i}\phi} \cr
-c_{d} s_{u}    & -s_{d}           & c_{d} c_{u} \cr} \right )
$
&& $\matrix{
(0.50^{+0.006}_{-0.016})^\circ+(27.75^{+8.734}_{-5.863})^\circ=(28.25^{+8.740}_{-5.879})^\circ \cr
(2.33^{+0.083}_{-0.085})^\circ+(34.49^{+4.169}_{-3.562})^\circ=(36.82^{+4.252}_{-3.647})^\circ \cr
(13.06^{+0.054}_{-0.060})^\circ+(42.43^{+6.631}_{-4.590})^\circ=(55.49^{+6.685}_{-4.650})^\circ \cr} $ \\ \\
%-----------------------------------------------------------------
\underline{{\it P9:} ~ $V \; = \; R_{31}(\rho) ~ R_{12}(\theta, \phi)
~ R_{23}(\sigma)$}
&& $\theta_u/\theta_d/\theta~~~~~~~~~~\vartheta_l/\vartheta_\nu/\vartheta~~~~~~~~~~~~~~~~~~~~~~  $
\\
$\left ( \matrix{
c^{~}_{\theta} c_{u}         & s^{~}_{\theta} c_{\sigma} c_{u} - s_{\sigma} s_{u} e^{-{\rm i}\phi}
& s^{~}_{\theta} s_{\sigma} c_{u} + c_{\sigma} s_{u} e^{-{\rm i}\phi} \cr
-s^{~}_{\theta} & c^{~}_{\theta} c_{\sigma}     & c^{~}_{\theta} s_{\sigma} \cr
-c^{~}_{\theta} s_{u}        & -s^{~}_{\theta} c_{\sigma} s_{u} - s_{\sigma} c_{u} e^{-{\rm i}\phi}
& -s^{~}_{\theta} s_{\sigma} s_{u} + c_{\sigma} c_{u} e^{-{\rm i}\phi} \cr} \right )
$
&& $\matrix{
(0.51^{+0.007}_{-0.016})^\circ+(24.86^{+5.223}_{-4.236})^\circ=(25.37^{+5.230}_{-4.252})^\circ \cr
(2.43^{+0.084}_{-0.088})^\circ+(48.17^{+8.282}_{-6.463})^\circ=(50.60^{+8.366}_{-6.551})^\circ \cr
(13.04^{+0.053}_{-0.059})^\circ+(24.09^{-2.114}_{+0.737})^\circ=(37.13^{-2.061}_{+0.678})^\circ \cr} $ \\ \\
\hline\hline
\end{tabular}
\end{center}
\label{9-Parameterizations}
\end{table}
\normalsize
%%%%%%%%%%%%%%%%%%%%%%%%%%%%%%%%%%%%%%%%%%%%%%%%%%%%%%%%%%%%%%%%%%%%%%%%%%%%


\begin{thebibliography}{99}\label{bib}


%\cite{Smirnov:2004ju}
\bibitem{SNO} SNO Collaboration, Q.R. Ahmad {\it et al.},
Phys. Rev. Lett. {\bf 89}, 011301 (2002).

\bibitem{SK} For a review, see: C.K. Jung {\it et al.},
Ann. Rev. Nucl. Part. Sci. {\bf 51}, 451 (2001).

\bibitem{KM} KamLAND Collaboration, K. Eguchi {\it et al.},
Phys. Rev. Lett. {\bf 90}, 021802 (2003).

\bibitem{K2K} K2K Collaboration, M.H. Ahn {\it et al.},
Phys. Rev. Lett. {\bf 90}, 041801 (2003).

\bibitem{King:2003jb}
see e.g.:%\cite{Mohapatra:2002kn}
  R.~N.~Mohapatra,
  %``ICTP lectures on theoretical aspects of neutrino masses and mixings,''
  arXiv:hep-ph/0211252;
  %%CITATION = HEP-PH/0211252;%%
  %for a review, see e.g.:
S.~F.~King,
%``Neutrino mass models,''
Rept.\ Prog.\ Phys.\  {\bf 67} (2004) 107, hep-ph/0310204;
%%CITATION = HEP-PH 0310204;%%
R.~N.~Mohapatra and P.~B.~Pal,
  %``Massive neutrinos in physics and astrophysics. Second edition,''
  [World Sci.\ Lect.\ Notes Phys.\  {\bf 72} (2004) 1];
  %%CITATION = 00327,60,1;%%
%\cite{Altarelli:2004za}
%\bibitem{Altarelli:2004za}
  G.~Altarelli and F.~Feruglio,
  %``Models of neutrino masses and mixings,''
  New J.\ Phys.\  {\bf 6} (2004) 106
  [arXiv:hep-ph/0405048];
  %%CITATION = NJOPF,6,106;%%
%\cite{Frampton:2004vw}

%\cite{Maki:1962mu}
\bibitem{MNS}
  Z.~Maki, M.~Nakagawa and S.~Sakata,
  %``Remarks on the unified model of elementary particles,''
  Prog.\ Theor.\ Phys.\  {\bf 28} (1962) 870; see also:
  %%CITATION = PTPKA,28,870;%%
      %\cite{Pontecorvo:1967fh}
    %  \bibitem{Pontecorvo:1967fh}
        B.~Pontecorvo,
        %``Neutrino experiments and the question of leptonic-charge  conservation,''
        Sov.\ Phys.\ JETP {\bf 26} (1968) 984
        [Zh.\ Eksp.\ Teor.\ Fiz.\  {\bf 53} (1967) 1717];
        %%CITATION = ZETFA,53,1717;%%
%\cite{Cabibbo:1963yz}
\bibitem{CKM}
  N.~Cabibbo,
  %``Unitary Symmetry and Leptonic Decays,''
  Phys.\ Rev.\ Lett.\  {\bf 10} (1963) 531;
  %%CITATION = PRLTA,10,531;%%
%\cite{Kobayashi:1973fv}
%\bibitem{Kobayashi:1973fv}
  M.~Kobayashi and T.~Maskawa,
  %``CP Violation In The Renormalizable Theory Of Weak Interaction,''
  Prog.\ Theor.\ Phys.\  {\bf 49} (1973) 652.
  %%CITATION = PTPKA,49,652;%%

\bibitem{Propose_smirnov}
  A.~Y.~Smirnov,
  %``Neutrinos: '...Annus mirabilis',''
  arXiv:hep-ph/0402264;
  %%CITATION = HEP-PH/0402264;%%%\cite{Raidal:2004iw}

\bibitem{Propose_Raidal}
  M.~Raidal,
  %``Relation between the neutrino and quark mixing angles and grand
  %unification,''
  Phys.\ Rev.\ Lett.\  {\bf 93} (2004) 161801
  [arXiv:hep-ph/0404046];
%\cite{Minakata:2004xt}
%\bibitem{Propose_Minakata}
  H.~Minakata and A.~Y.~Smirnov,
  %``Neutrino Mixing and Quark-Lepton Complementarity,''
  Phys.\ Rev.\  D {\bf 70} (2004) 073009
  [arXiv:hep-ph/0405088];
  %%CITATION = PHRVA,D70,073009;%%
  %%CITATION = PRLTA,93,161801;%%
 %\bibitem{Frampton:2004vw}
  P.~H.~Frampton and R.~N.~Mohapatra,
  %``Possible gauge theoretic origin for quark-lepton complementarity,''
  JHEP {\bf 0501} (2005) 025
  [arXiv:hep-ph/0407139];
  %%CITATION = JHEPA,0501,025;%%
  %\cite{Xing:2005ur}
%\bibitem{Xing:2005QLC}
  Z.~z.~Xing,
  %``Nontrivial correlation between the CKM and MNS matrices,''
  Phys.\ Lett.\  B {\bf 618} (2005) 141
  [arXiv:hep-ph/0503200];
  %%CITATION = PHLTA,B618,141;%%
  %\cite{Antusch:2005ca}
%\bibitem{Antusch:2005ca}
  S.~Antusch, S.~F.~King and R.~N.~Mohapatra,
  %``Quark lepton complementarity in unified theories,''
  Phys.\ Lett.\  B {\bf 618} (2005) 150
  [arXiv:hep-ph/0504007];
  %%CITATION = PHLTA,B618,150;%%
 %\cite{Falcone:2005sa}
%\bibitem{Falcone:2005sa}
  D.~Falcone,
  %``Quark lepton symmetry and complementarity,''
  Mod.\ Phys.\ Lett.\  A {\bf 21} (2006) 1815
  [arXiv:hep-ph/0509028].
  %%CITATION = MPLAE,A21,1815;%%
%\cite{Minakata:2005rf}


%\cite{Hochmuth:2006xn}
\bibitem{phenomenology}
  K.~A.~Hochmuth and W.~Rodejohann,
  %``Low and High Energy Phenomenology of Quark-Lepton Complementarity
  %Scenarios,''
  Phys.\ Rev.\  D {\bf 75} (2007) 073001
  [arXiv:hep-ph/0607103];
  %%CITATION = PHRVA,D75,073001;%%
%\cite{Picariello:2007ss}
%\bibitem{Picariello:2007ss}
  M.~Picariello, B.~C.~Chauhan, J.~Pulido and E.~Torrente-Lujan,
  %``Predictions from non trivial Quark-Lepton complementarity,''
  Int.\ J.\ Mod.\ Phys.\  A {\bf 22} (2008) 5860
  [arXiv:0706.2332 [hep-ph]].
  %%CITATION = IMPAE,A22,5860;%%

%\cite{Schmidt:2006rb}
\bibitem{Smirnov_RGE}

  S.~K.~Kang, C.~S.~Kim and J.~Lee,
  %``Quark-lepton complementarity with renormalization effects through threshold
  %corrections,''
  PoS {\bf HEP2005} (2006) 189
  [arXiv:hep-ph/0512023];
  %%CITATION = POSCI,HEP2005,189;%%
  M.~A.~Schmidt and A.~Y.~Smirnov,
  %``Quark lepton complementarity and renormalization group effects,''
  Phys.\ Rev.\  D {\bf 74} (2006) 113003
  [arXiv:hep-ph/0607232];
  %%CITATION = PHRVA,D74,113003;%%
%\cite{Dighe:2007ksa}
%\bibitem{Dighe:2007ksa}
  A.~Dighe, S.~Goswami and P.~Roy,
  %``Radiatively broken symmetries of nonhierarchical neutrinos,''
  Phys.\ Rev.\  D {\bf 76} (2007) 096005
  [arXiv:0704.3735 [hep-ph]].
  %%CITATION = PHRVA,D76,096005;%%


 %\cite{Plentinger:2006nb}
\bibitem{extended_QLC}
  F.~Plentinger, G.~Seidl and W.~Winter,
  %``Systematic parameter space search of extended quark-lepton
  %complementarity,''
  Nucl.\ Phys.\  B {\bf 791} (2008) 60
  [arXiv:hep-ph/0612169];
  %%CITATION = NUPHA,B791,60;%%
%\cite{Plentinger:2007px}
%\bibitem{Plentinger:2007px}
  F.~Plentinger, G.~Seidl and W.~Winter,
  %``The Seesaw Mechanism in Quark-Lepton Complementarity,''
  Phys.\ Rev.\  D {\bf 76} (2007) 113003
  [arXiv:0707.2379 [hep-ph]].
  %%CITATION = PHRVA,D76,113003;%%

\bibitem{Review}
  H.~Minakata,
  %``Quark-lepton complementarity: A review,''
  arXiv:hep-ph/0505262;
  %%CITATION = HEP-PH/0505262;%%
%\cite{Jarlskog:2005jn}
%\cite{GonzalezCanales:2006gd}
%\bibitem{GonzalezCanales:2006gd}
  F.~Gonzalez Canales and A.~Mondragon,
  %``On quark-lepton complementarity,''
  AIP Conf.\ Proc.\  {\bf 857} (2006) 287
  [arXiv:hep-ph/0606175].
  %%CITATION = APCPC,857,287;%%

\bibitem{Jarlskog}
  C.~Jarlskog,
  %``Ambiguities pertaining to quark-lepton complementarity,''
  Phys.\ Lett.\  B {\bf 625} (2005) 63
  [arXiv:hep-ph/0507212].
  %%CITATION = PHLTA,B625,63;%%

%\cite{Fritzsch:1997st}
\bibitem{FXpara}
%\cite{Fritzsch:1997fw}
%\bibitem{Fritzsch:1997fw}
  H.~Fritzsch and Z.~Z.~Xing,
  %``Flavor symmetries and the description of flavor mixing,''
  Phys.\ Lett.\  B {\bf 413} (1997) 396
  [arXiv:hep-ph/9707215];
  %%CITATION = PHLTA,B413,396;%%
  H.~Fritzsch and Z.~z.~Xing,
  %``On the parametrization of flavor mixing in the standard model,''
  Phys.\ Rev.\  D {\bf 57} (1998) 594
  [arXiv:hep-ph/9708366].
  %%CITATION = PHRVA,D57,594;%%

\bibitem{Fritzsch-Xing}
  H.~Fritzsch and Z.~z.~Xing,
  %``Mass and flavor mixing schemes of quarks and leptons,''
  Prog.\ Part.\ Nucl.\ Phys.\  {\bf 45} (2000) 1
  %[arXiv:hep-ph/9912358].

  %\cite{Amsler:2008zzb}
\bibitem{PDG}
  C.~Amsler {\it et al.}  [Particle Data Group],
  %``Review of particle physics,''
  Phys.\ Lett.\  B {\bf 667} (2008) 1.
  %%CITATION = PHLTA,B667,1;%%


%\cite{Strumia:2006db}
\bibitem{Strumia}
  A.~Strumia and F.~Vissani,
  %``Neutrino masses and mixings and.,''
  arXiv:hep-ph/0606054.
  %%CITATION = HEP-PH/0606054;%%


%\cite{Xing:1994rj}
\bibitem{Wolfenstein}
  Z.~z.~Xing,
  %``Wolfenstein parametrization revisited,''
  Phys.\ Rev.\  D {\bf 51} (1995) 3958;
  %[arXiv:hep-ph/9411340].
  %%CITATION = PHRVA,D51,3958;%%
 %\bibitem{Wolfenstein_Paganini}
  %\cite{Paganini:1997cu}
  P.~Paganini, F.~Parodi, P.~Roudeau and A.~Stocchi,
  %``Measurements of the $\rho$ and $\eta$ parameters of the V(CKM) matrix and
  %perspectives,''
  Phys.\ Scripta {\bf 58} (1998) 556;
 % [arXiv:hep-ph/9711261].
  %%CITATION = PHSTB,58,556;%%
  %\bibitem{Wolfenstein_CKMfitter}
J.~Charles {\it et al.}  [CKMfitter Group],
  %``CP violation and the CKM matrix: Assessing the impact of the asymmetric $B$
  %factories,''
  Eur.\ Phys.\ J.\  C {\bf 41} (2005) 1.
%  [arXiv:hep-ph/0406184].
  %%CITATION = EPHJA,C41,1;%%



%\cite{Fogli:2009zza}
\bibitem{Fogli:2009zza}
 see e.g. G.~L.~Fogli, E.~Lisi, A.~Marrone, A.~Palazzo and A.~M.~Rotunno,
  %``Neutrino Masses And Mixing: 2008 Status,''
  Nucl.\ Phys.\ Proc.\ Suppl.\  {\bf 188} (2009) 27;
  %%CITATION = NUPHZ,188,27;%%
  %\cite{Schwetz:2008er}
%\bibitem{Schwetz:2008er}
  T.~Schwetz, M.~A.~Tortola and J.~W.~F.~Valle,
  %``Three-flavour neutrino oscillation update,''
  New J.\ Phys.\  {\bf 10} (2008) 113011
  [arXiv:0808.2016 [hep-ph]].
  %%CITATION = NJOPF,10,113011;%%

%\cite{Fogli:2005cq}
%\bibitem{Fogli:0506083}
 % G.~L.~Fogli, E.~Lisi, A.~Marrone and A.~Palazzo,
  %``Global analysis of three-flavor neutrino masses and mixings,''
  %Prog.\ Part.\ Nucl.\ Phys.\  {\bf 57} (2006) 742
  %[arXiv:hep-ph/0506083].
  %%CITATION = PPNPD,57,742;%%

%\cite{Antusch:2003kp}
\bibitem{Antusch:2003kp}
  S.~Antusch, J.~Kersten, M.~Lindner and M.~Ratz,
  %``Running neutrino masses, mixings and CP phases: Analytical results and
  %phenomenological consequences,''
  Nucl.\ Phys.\  B {\bf 674}, 401 (2003)
  [arXiv:hep-ph/0305273].
  %%CITATION = NUPHA,B674,401;%%
 %\cite{Kang:2005ck}
%\bibitem{Kang:2005ck}
  \bibitem{XRGE_q}
  Z.~z.~Xing,
  %``Right Unitarity Triangles, Stable CP-violating Phases and Approximate
  %Quark-Lepton Complementarity,''
  Phys.\ Lett.\  B {\bf 679} (2009) 111
  %[arXiv:0904.3172 [hep-ph]].
  %%CITATION = PHLTA,B679,111;%%
%\cite{Fogli:2008ig}

%\cite{Xing:2005fw}
\bibitem{XRGE_l}
  Z.~z.~Xing,
  %``A novel parametrization of tau-lepton dominance and simplified one-loop
  %renormalization-group equations of neutrino mixing angles and  CP-violating
  %phases,''
  Phys.\ Lett.\  B {\bf 633} (2006) 550
  %[arXiv:hep-ph/0510312].
  %%CITATION = PHLTA,B633,550;%%
%\cite{Xing:2009eg}

 %\cite{Xing:2007fb}
\bibitem{Xing_running}
 Z.~z.~Xing, H.~Zhang and S.~Zhou,
  %``Updated Values of Running Quark and Lepton Masses,''
  Phys.\ Rev.\  D {\bf 77} (2008) 113016
  [arXiv:0712.1419 [hep-ph]].
  %%CITATION = PHRVA,D77,113016;%%


%\cite{Komatsu:2008hk}
%\bibitem{WMAP_mass}
 % E.~Komatsu {\it et al.}  [WMAP Collaboration],
  %``Five-Year Wilkinson Microwave Anisotropy Probe (WMAP\altaffilmark 1 )
  %Observations:Cosmological Interpretation,''
  %Astrophys.\ J.\ Suppl.\  {\bf 180} (2009) 330
  %[arXiv:0803.0547 [astro-ph]].
  %%CITATION = APJSA,180,330;%%

%\bibitem{Dimopoulos}
 % S.~Dimopoulos, L.~J.~Hall and S.~Raby,
  %``A Predictive ansatz for fermion mass matrices in supersymmetric grand
  %unified theories,''
  %Phys.\ Rev.\  D {\bf 45} (1992) 4192.
  %%CITATION = PHRVA,D45,4192;%%


\end{thebibliography}
\end{document}